# Antenna enhanced infrared photoinduced force imaging in aqueous environment with super-resolution and hypersensitivity

Jian Li,[1] Jie Pang,[1] Zhen-dong Yan,[2] Junghoon Jahng,[3] Jin Li,[1] William Morrison,[4] Jing Liang,[1] Qing-Ying Zhang,[1] Xing-Hua Xia[1]*

1. State key laboratory of analytical chemistry for life science, school of chemistry and chemical engineering, Nanjing University, Nanjing 210023
2. College of science, Nanjing Forestry University, Nanjing 210037
3. Center for nanocharacterization, Korea Research Institute of Standards and Science, Daejeon 34113
4. Molecular Vista Inc., 6840 Via Del Oro, Suite 110, San Jose, CA 95119

*Corresponding Author(s): Xing-Hua Xia: xhxia@nju.edu.cn

**Abstract**

Tip enhanced IR spectra and imaging have been widely used in cutting-edge studies for the in-depth understanding of the composition, structure and function of interfaces at the nanoscale. However, molecular monolayer sensitivity has only been demonstrated at solid/gas interfaces. In aqueous environment, the reduced sensitivity limits the practical applications of tip enhanced IR nanospectroscopy. Here, we demonstrate an approach to hypersensitive nanoscale IR spectra and imaging in aqueous environment with the combination of photoinduced force (PiF) microscopy and resonant antennas. The highly confined electric field inbetween the tip and antenna amplifies the PiF, while the excitation *via* plasmon internal reflection mode minimizes the environmental absorption. A polydimethylsiloxane (PDMS) layer (~1-2 nm thickness) functionalized on the AFM tip and self-assembled monolayer of bovine hemoglobin on antenna has been successfully identified in water with antennas of different sizes. Sampling volume of ~604 chemical bonds from PDMS was demonstrated with sub-10 nm spatial resolution confirmed by electric field distribution mapping on antennas. This platform demonstrates for the first time the application of PiF microscopy in aqueous environments, providing a brand-new configuration to achieve highly enhanced nanoscale IR signals, which is promising for future research of interfaces and nanosystems in aqueous environments.

**Introduction**

Mid-infrared (mid-IR) spectroscopy has attracted extensive attentions for interfacial studies since IR absorption of molecules arises from specific chemical bonds, allowing the analysis of molecular structures and functions without labelling or destruction.[1,2] Recently, atomic force microscopy (AFM) based opto-mechanical force detection techniques such as photothermal induced resonance (PTIR),[3-6] photoinduced force microscopy (PiFM)[7-10] and peak force infrared (PFIR)[11-13] microscopy have been demonstrated to achieve IR analysis with nanoscale spatial resolution, allowing subwavelength scale investigation of chemical and physical properties of molecules. At solid/gas interfaces, the detection sensitivity has hinted at molecular monolayer level,[3,7] which is promising to reveal information about the mechanisms of interfacial processes. However, challenges arise when applying opto-mechanical force based IR techniques in aqueous environment: (1) Decrease of the sensitivity due to the damping of cantilever oscillation in water; (2) Strong background water absorption of the incident light.[14] These intrinsic problems severely limit the range of application of AFM based IR techniques. Since aqueous environment could sustain the natural structure and function of biological molecules and also provide an environment for bio/electro-chemical reactions, the development of nanoscale IR techniques with interfacial molecular (mono)layer level sensitivity in aqueous environment is greatly desirable.

Pioneering works have been demonstrated on PTIR and PFIR by introducing attenuated total reflection (ATR) configurations,[14-17] which concentrate the incident light from the substrate side and enhance the electric field near the prism surface by one order of magnitude, allowing detection of polymer thin film with thicknesses as small as 20 nm. However, in the previous works, the bulk materials entirely contribute to the opto-mechanical signals, while in interfacial studies, molecular monolayer level sensitivity is usually required for the analysis of heterogeneous processes. Thus, groundbreaking design is yet highly desired for practical interfacial IR analysis in aqueous environment.

An antenna is a well-defined structure that can effectively confine the electric field on hot spot when it is resonant with the incident light of a specific wavelength.[18] Within the hot spot, IR absorption of molecules is considerably increased, leading to a greatly enhanced IR response that couples with the antenna's optical response. Using back side excitation *via* "plasmon internal reflection" mode[19-21] or ATR[22] mode, antenna arrays have already been used for IR analysis in the aqueous environment with high sensitivity and background absorption suppression. Thus, a combination of back side antenna excitation and the opto-mechanical force based IR techniques could be a promising means to improve the sensitivity of nanoscale IR analysis in aqueous environment.

Photoinduced force (PiF) originates from both the dipole force (photoinduced electromagnetic force)[23] and the thermal expansion-modulated Van der Waals force (photoinduced thermodynamic force)[24] of molecules close to the tip, which are both more sensitive to interfacial molecules than bulk materials. In this study, we demonstrate that hypersensitive nanoscale IR analysis (of ~1-2 nm thickness of polydimethylsiloxane, PDMS, and monolayer bovine hemoglobin) can be realized *via* antenna enhanced PiFM. The antenna-enhanced PiF spectra follow the dissipation line shape, exhibiting the feasibility for interfacial chemical identification. With the electric field dependent molecular signal, the hot spot on the antenna is mapped with spatial resolution up to sub-10 nm, suggesting a ~77 yoctoliter sampling volume with ~604 chemical bonds detection limit. In addition, the PiF spectral shape recorded on antennas exhibits strong wavelength dependence, emphasizing the importance of tunable antenna resonance for wide spectral range nanoscale IR analysis in aqueous environment.

**Experimental Section**

**Material**

Polystyrene (PS) microspheres (wt.:1%) with different diameters of 4, 5, and 6 μm were bought from Suzhou Nanomicro Technology Company, Ltd (China). ZnSe wafer (diameter: 25 mm, thickness: 0.5 mm) was bought from Qingxuan Technology (Changchun, China). AFM tip (NCH-PtIr-PiFM-50) was bought from Molecular Vista, Inc. (San Jose, USA). Bovine hemoglobin (BHb) was from Biological Products Corp. (Shanghai, China) and used without further purification. Other reagents and chemicals were of analytical grade. All reagents were used as received without further purification. All solutions were prepared with Milli-Q water from a Millipore system.

**Instruments**

Atomic force microscopic (AFM) and photoinduced force microscopic (PiFM) images were recorded on a VistaScope system (Molecular Vista, USA). Fourier transform infrared (FTIR) spectra were collected on a Nicolet IS50 (Nicolet, USA). Vacuum evaporation of Au was carried out on a PVD75 Proline SP instrument (Kurt, J Lesker, USA).

**Fabrication of antenna arrays on ZnSe wafer**

Self-assembled monolayer (SAM) mask of PS spheres covered substrate was fabricated by a mature Langmuir–Blodgett method developed by our group.[22] The fabrication of antenna array was carried out by setting the PS SAM mask covered substrate into a sputtering instrument with the mask facing to Au target. Vacuum evaporation of gold film (50 nm thickness) on the substrate was performed with 1 Angstrom/s speed. After the PS SAM mask was removed by scotch tape, the remained triangle gold antenna array on the substrate was cleaned by deionized water.

**Measurement of the optical properties of the triangle gold antenna arrays**

Transmission spectra of the triangle gold antenna arrays in air were acquired using the spectrum of pure ZnSe wafer in air as the reference. The reflectance spectra of the triangle gold antenna arrays in water were acquired using the reflectance spectrum of ZnSe wafer coated with Au film (thickness: 100 nm) in water as the reference.

**Fabrication of bovine hemoglobin monolayer on antenna**

The monolayer of bovine hemoglobin was assembled from an aqueous solution of 1 mg/mL BHb covered on the top of a freshly fabricated antenna surface for 1 h. Then, the sample was washed by deionized water for 3 times and dried by nitrogen gas.

**Measurement of PiFM**

PiFM measurement was taken on a VistaScope microscope that is coupled to a quantum cascade laser (QCL) laser system from Block Engineering with a wavenumber resolution of 1 cm$^{-1}$ and a tuning range from 770 to 1885 cm$^{-1}$. For detection, the IR beam was focused with an objective lens (×52/NA: 0.6, working distance: 1.6 mm) from the bottom side of ZnSe substrate. The IR beam has a pulse duration of 44 ns and pulse energy of 0.1 μJ–2 μJ with incident wavelengths. The set point for detection was set as 75% with an oscillation amplitude of ~1 nm (in air) and ~3 nm (in water). The microscope was operated in dynamic mode with NCH-PtIr-PiFM-50 non-contact cantilevers from Molecular Vista, Inc. The cantilever was excited at its second resonance around 1650 KHz (in air) and 876 KHz (in water). The collection time for each spectrum was around 1 s, and the time per image was about 2 min and 4 min with 1 line/s speed at 128×128 and 256×256 resolution, respectively. The spectrum was normalized with the background laser power profile spectrum.

**Simulations**

Numerical simulations were performed using a finite-element-method-based software package (COMSOL Multiphysics). All simulations were performed by scattering field method. The refractive index of ZnSe wafer, water and air were taken as 2.4, 1.33 and 1, respectively. The permittivity of gold was described by a Lorenz-Drude model. The permittivity of PDMS was acquired from Ref. 25. To qualitatively simulate the absorption of PDMS, the tip was set as a hemisphere with 10 nm radius head connected a cone with 30 nm end radius and 200 nm height, which was 2 nm above the antenna tip end. The PDMS was simulated as a cylinder with 5 nm radius and 1 nm height on top of the antenna tip end.

**Results and Discussion**

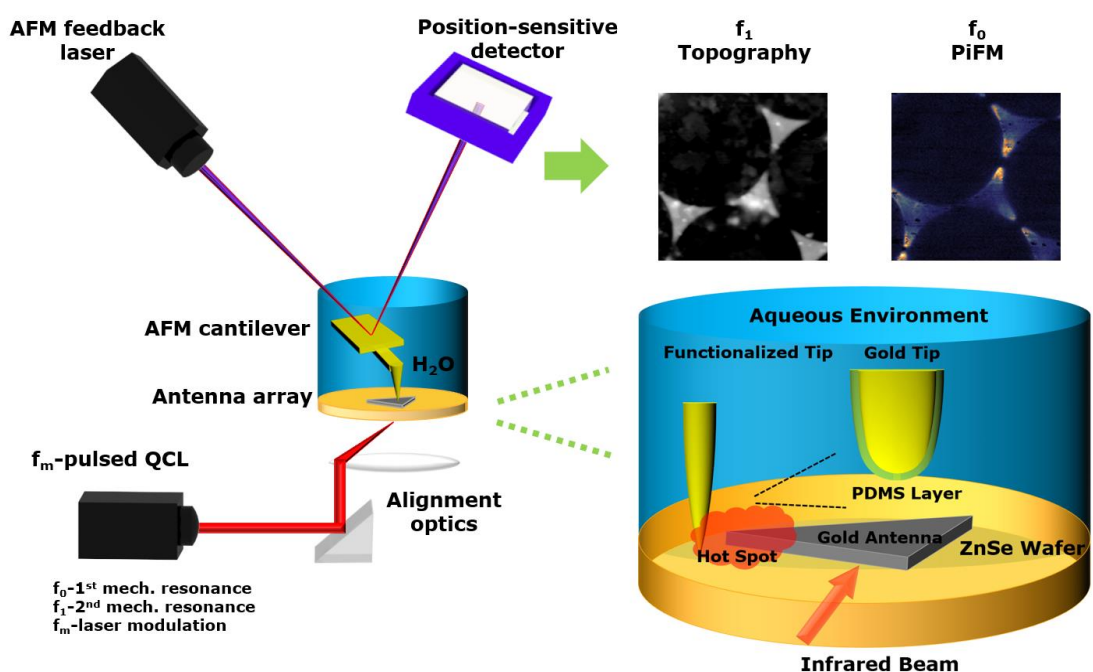

**Scheme 1.** Schematic illustration of the antenna enhanced PiFM detection platform in aqueous environment.

The antenna enhanced PiFM detection platform in aqueous environment is illustrated in Scheme 1. Wafer scale Au triangle antenna array are fabricated on top of the visible to mid-IR transparent ZnSe substrate (0.5 mm thickness). An objective lens with long working distance is used to focus the non-polarized incident light from a quantum cascade laser (pulse: 44 ns). The backside illumination can minimize the absorption of aqueous solution and efficiently excite the antenna resonance. The difference ($f_m$) between the first ($f_0$) and second ($f_1$) oscillation frequencies of the AFM tip cantilever, defined as side band modulation mode, is applied as the repetition rate of the laser since it is demonstrated more sensitive to the force gradient. In the tapping mode detection, $f_0$ is used to detect the PiF signal, while $f_1$ is used to detect the topography. PiFM detection at $f_m$ would introduce a modulation of the tip–sample gap distance as well as the tip–sample interaction force, leading to the sideband coupling through the derivative of the tip–sample interaction. Thus, in the sideband measurement, for a given molecular resonance with specific thickness $d$ of an organic monolayer, the photoinduced thermodynamic force, the thermal expansion-modulated Van der Waals force (ΔF$_{vdW}$), is dominant to the photoinduced electromagnetic force (dipole force),[26] given as:

$$\Delta F_{vdW} \propto \frac{\partial F_{vdW}}{\partial z}\sigma d \int a_{abs}|E|^2 dz \tag{1}$$

where, F$_{vdW}$ is the Van der Waals force, which is the typical tip–sample interaction force, $\sigma$ is the linear thermal expansion coefficient, $a_{abs}$ is the absorption coefficient of organic layer and $E$ is the strength of electric field. With the near field opto-mechanical force transformation, PiF signal of the molecules inbetween tip and antenna can be recorded with high enhancement (Figure 1).

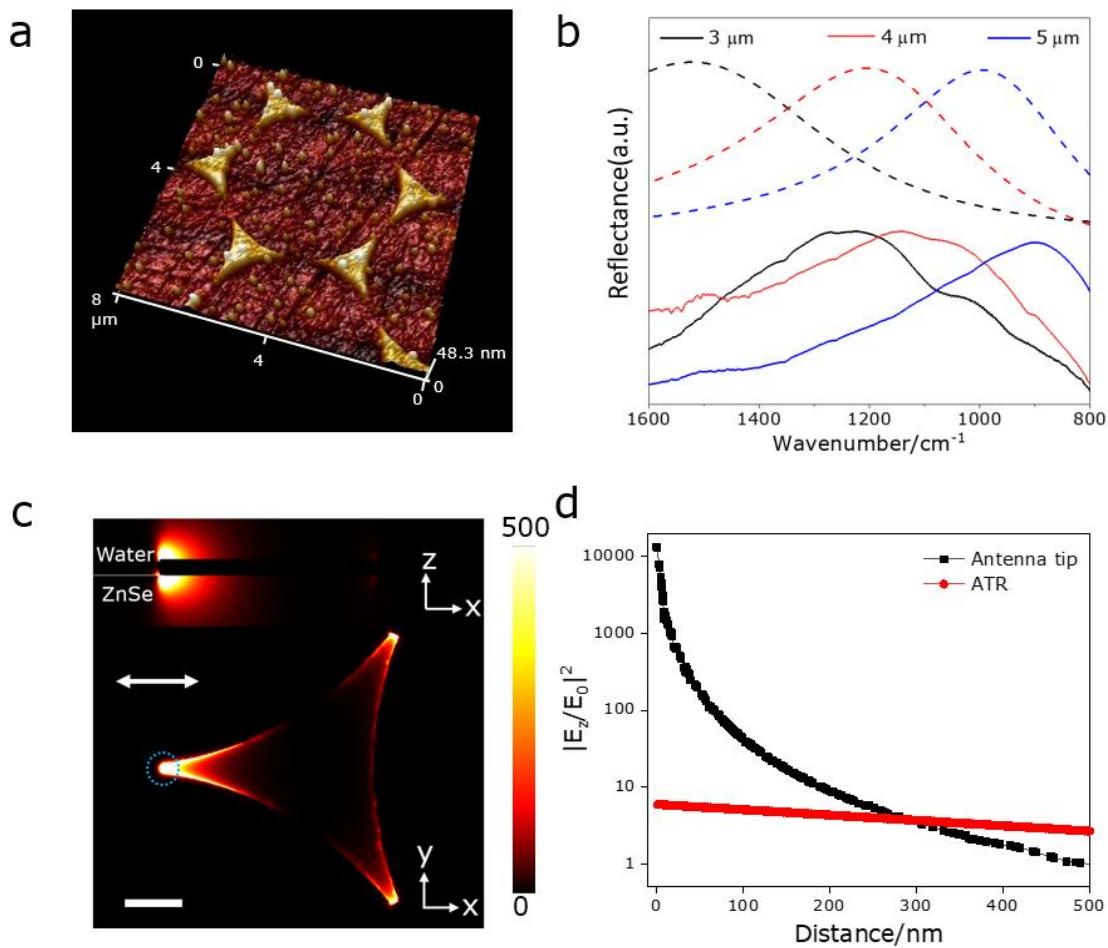

**Figure 1. Properties of the fabricated gold triangle antenna arrays**. (a) AFM image of an antenna array fabricated with a mask of 5 μm PS sphere. (b) Measured (solid curves) and simulated resonance spectra (normalized, dotted curves) in water of the antenna arrays fabricated with masks of 4, 5 and 6 μm PS spheres, respectively. (Substrate: ZnSe: n=2.4; incident angle: 12º). (c) Simulated $|E_z/E_0|^2$ distribution on an antenna fabricated with a mask of 5 μm PS sphere. (Wavenumber: 1205 cm$^{-1}$; xy plane: 1 nm away in water from the antenna surface, xz plane: along the main axis of the antenna; the incident light is with *s* polarization as indicated by the double-head arrow with incident angle of 12°; scale bar in xy panel: 400 nm). (d) E$_z$ field intensity in the sampling region with refractive index of water (n=1.33). Black curve: at the antenna tip as pointed by a dashed blue circle in (c); Red curve: calculation for *p* polarization on ZnSe with an incident angle of 45°. Wavenumber: 1205 cm$^{-1}$.

Figure 1a shows the AFM topography of an antenna array fabricated on a ZnSe substrate *via* nanosphere lithography with a mask of 5 μm PS sphere. The highly ordered triangle antennas with around 50 nm height are separated with more than 100 nm distance to avoid coupling. Utilization of triangle array on flat wafer has two main advantages: (1) Triangle antennas are demonstrated to have anti-polarization properties.[22] Thus, non-polarized incident light can be used, which simplifies the optical alignment and avoids power reduction from a polarizer. (2) Focus of the incident beam is more convenient than the evanescent wave achieved using an ATR prism. The detection region can be located at an arbitrary selected region on the whole wafer.

Figure 1b exhibits the measured and simulated antenna resonance spectra of antenna arrays with different sizes in water (backside excitation with 12° of incident angle). For simplification, the simulation only using *s*-polarization along the main axis on single antenna is shown in Figure 1c. As is clearly seen, the simulated antenna resonance correlates well with the experimental data, the slight shift of peak position could be owing to the shape and size variations of the real antenna with the model. As compared to the antenna resonance in air, the resonance of the antenna with the same size in water red-shifts around 100~200 cm$^{-1}$ (Figure S1) because of the larger refractive index of water (n=1.33) than air (n=1). By changing the PS sphere from 4 μm to 6 μm, resonance of the resulted antennas shifts from ~1205 cm$^{-1}$ to ~900 cm$^{-1}$. In addition, we find that the incident angle has minute influence on the resonance peak position as revealed by simulation (Figure S2), suggesting that the effect of different incident angle caused by objective lens can be minimized.

The better confinement of electric field makes the antenna array a highly advantageous alternative to an ATR configuration. As the simulations in Figure 1c and Figure S3 show, when the plasmon resonance is excited by the light incident from the substrate side, enhanced electric field is generated at the tip end of the antenna in the aqueous environment, providing unique opportunities to increase the light matter interaction in such a hot spot region. The electric field enhancement is also compared with the normally used ATR configuration (Figure 1d) in the sampling region. For ATR excitation with 45° *p*-polarized light illumination, the intrinsic $E_z$ field enhancement is only 5.9, while on the antenna tip, the enhancement is more than 1000 times of that on the ATR prism surface. Since PiF is proportional to $|E|^2$, the PiF signal on antenna is expected to be a couple of orders of magnitude higher than on ATR prism. In addition, the electric field is only enhanced close to the antenna surface within less than 50 nm, while in ATR configuration the field would penetrate more than 500 nm distance, suggesting that the near field sensitivity of antenna is much higher than ATR. These advantages make the antenna array a better platform for highly sensitive PiFM measurements, especially for interfacial studies.

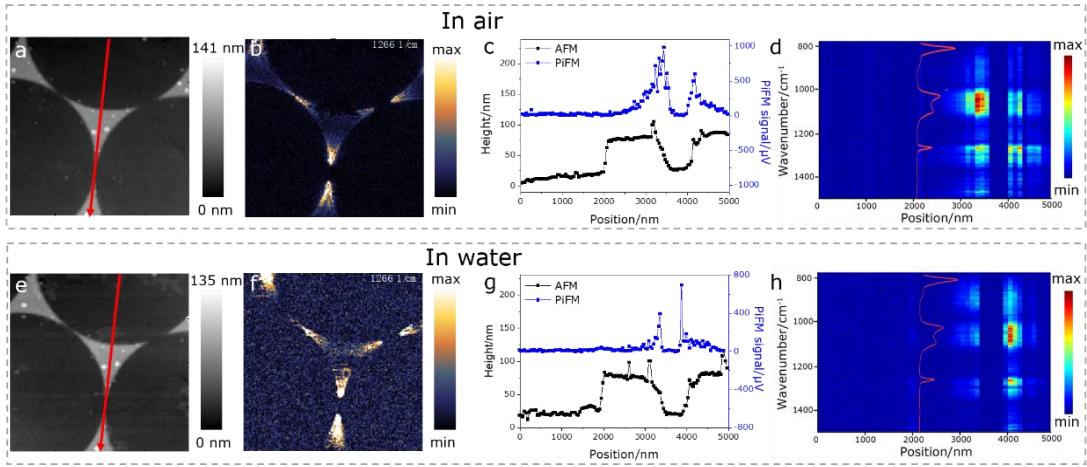

**Figure 2. PiFM measurements on an antenna fabricated with a mask of 5 μm PS sphere.** (a) and (e): AFM images of the same region recorded in air and water, respectively. Scale bar: 1 μm. (b) and (f): PiF images of the same region of the antennas in (a) and (e) simultaneously recorded at 1266 cm$^{-1}$, respectively. (c) and (g): Signals along the red arrows of the topography (black curves) and PiFM images (blue curves). (d) and (i): Corresponding spatio-spectral transection along the red arrows. Spectra were obtained in 5 s with 1 cm$^{-1}$ spectral resolution, with 100 nm separation.

To demonstrate the possibility of the proposed strategy, the PDMS layer on a commercial PtIr AFM tip is firstly used as probe, which is naturally coated on the tip in the packaging box with the thickness ~1–2 nm.[27,28] Such sampling strategy has several advantages: (1) Sample variation on different detection region could be eliminated, making it reliable to compare the signal enhancement at different positions and different antennas; (2) PDMS has four characteristic IR peaks at 1266 cm$^{-1}$, 1100 cm$^{-1}$, 1025 cm$^{-1}$ and 816 cm$^{-1}$ (dielectric function shown in Figure S4), which overlap well with the resonance of the applied antenna arrays; (3) PDMS has relatively high absorption coefficient and large thermal expansion coefficient which are expected to induce strong photoinduced thermodynamic force.[26]

We first performed PiF studies on the antenna array fabricated with 5 μm PS sphere in air and water in the same region. The $f_0$ and $f_1$ frequencies of the cantilever shift from 262.2 kHz and 1653 kHz to 128.2 kHz and 875.8 kHz, respectively, due to the oscillation damping of the cantilever and in water. In addition, the water environment will significantly decrease the Q factor of the cantilever. The exact Q decrease is difficult to be estimated because once in liquid, the resonance peak used becomes too asymmetric and unpredictable to effectively measure a full width at half maximum. Figure 2a and 2e show the AFM images of the same region recorded in air and water, respectively. The similar geometry and defects suggest that the detection region are on the same antenna. As shown in Figure 2b and 2f, the simultaneously recorded PiF signals at 1266 cm$^{-1}$ clearly exhibit position dependent optical responses. Since the PDMS layer is modified on AFM tip, the artifact due to the variation in molecules amount can be excluded, the observed signal can be concluded from the electric field induced light/matter interaction at the different position of the antenna. Meanwhile, the expansion of PDMS layer on the whole tip cone within illumination region has negligible effect since the PiF signal on top of ZnSe is only at the noise level. According to formula (1), the measured PiF signal mainly arises from the Van der Waals force induced by the PDMS expansion, which is proportional to the localized electric field intensity. Thus, despite the tapping mode operation induced gap size variation and heat diffusion in water, the collected PiF image can be qualitatively used to image the electric field distribution on the antenna. As demonstrated at different wavenumbers (Figure 2f and S5), the antenna tip end has the highest PiF signal, which is also in agreement with the simulated electric filed distribution as shown in Figure 1c.

We further collected the PiF spectra of PDMS on the antenna along the red arrows indicated in Figure 2a and 2e, the relative AFM and PiF signals in images were extracted and plotted in Figure 2c and 2g for comparison, respectively. The position dependent PiF spectra are shown in Figure 2d and 2h with the same position coordinate of Figure 2c and 2f. The PiF spectra of the tip clearly show the molecular features of PDMS at 1266 cm$^{-1}$, 1100 cm$^{-1}$ and 1020 cm$^{-1}$, which are in agreement with the far field band characteristics (red line in Figure 2d and 2h). Because the PDMS layer is expected as 1-2 nm thick, the hypersensitivity in aqueous environment can be achieved in our platform, which is an important step forward for nanoscale IR analysis. The observed spectra follow the dissipative lineshape (photoinduced thermodynamic force) rather than the dispersive lineshape (photoinduced electromagnetic force) of PDMS or the antenna resonance,[23] suggesting that the absorption induced Van der Waals force modulation is dominant inbetween the tip-PDMS-antenna junction. These distinct spectral features suggest that the proposed method can be used for the nanoscale chemical identification. In addition, the PiF spectral signal is strong near the tip end of the antenna and decrease rapidly when the detection position moves to the antenna surface, which is in agreement with the recorded PiF images and extracted $E_z$ strength in the simulation (Figure S6), confirming the origin of the PiF images to the antenna enhanced thermal expansion of target molecules.

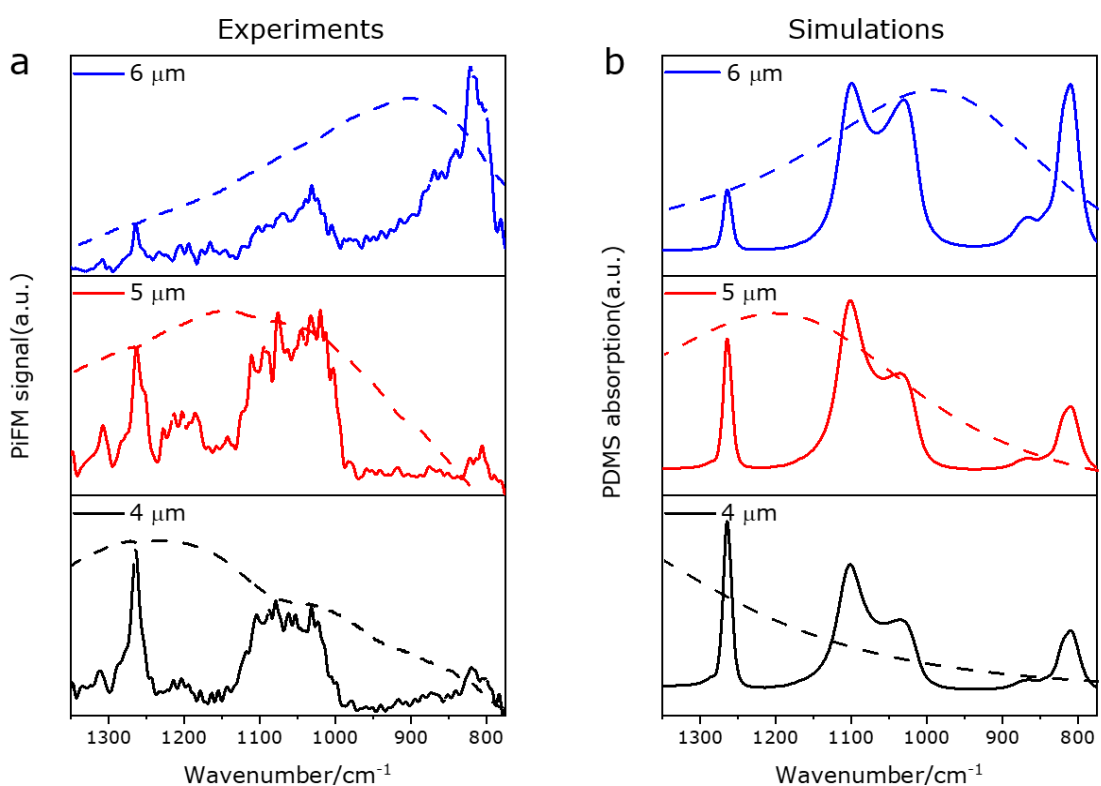

**Figure 3. PiF spectra on the antennas fabricated with PS spheres of different sizes.** (a) Measured PiFM spectra and (b) simulated PDMS absorption profiles at the tip end of the antennas fabricated with PS spheres of different sizes. Dashed lines show the relative antenna resonance profiles for comparison.

We then performed the nanoscale IR spectra and imaging on the antenna arrays fabricated with PS sphere diameters of 4 μm and 6 μm. The acquired AFM and PiFM images are of high quality and the IR spectra are successfully recorded with position (Figure S7-S9), demonstrating the great feasibility and reproducibility of our antenna enhanced PiFM approach for the in aqueous nanoscale IR analysis with hypersensitivity. Since PDMS has 4 characteristic peaks within the resonance region of the antenna, the PiFM results provide a unique way to control the coupling between different molecular vibrations and the antenna resonances. Figure 3(a) exhibits the PiFM spectra of PDMS measured at the tip end of the antennas fabricated with different sized PS spheres. On the antenna fabricated with 4 μm PS sphere with resonance at 1205 cm$^{-1}$, the peak at 1266 cm$^{-1}$ is much higher than at 816 cm$^{-1}$, which is reversed on antenna fabricated with 6 μm PS spheres with resonance at 898 cm$^{-1}$ that the peak at 1266 cm$^{-1}$ is much lower than at 816 cm$^{-1}$. This can be well understood that on the same antenna, higher absorption of the molecule is expected when its vibration coupling to the antenna resonance. Although this phenomenon is widely accepted and exhibited in the far field experiments, our work is the first near field demonstration in aqueous environment for the resonance dependent peak enhancement of molecules on the same antenna. Meanwhile, the enhancement of the same peak of PDMS with different antenna resonances is longitudinally compared. As the antenna size increases, the antenna resonance peak red-shifts from 1205 cm$^{-1}$ to 898 cm$^{-1}$. The peak strength at 1266 cm$^{-1}$ decreases due to the gradually mismatch with the antenna resonance, while the peak at 816 cm$^{-1}$ increases since it couples better with the longer wavelength resonance. For the adjacent peaks at 1100 cm$^{-1}$ and 1020 cm$^{-1}$, the relative strength of the PiF signals exhibit the similar tendency. These results are supported by the simulation on PDMS absorption underneath the AFM tip (details refer to the experimental section), demonstrating the effective coupling between antenna, tip and PDMS molecules. For comparison, we performed PiFM measurement on a clean ZnSe substrate in water, and did not observe any PiFM signal, which further emphasizes the importance of antenna resonance for sensitive detection.

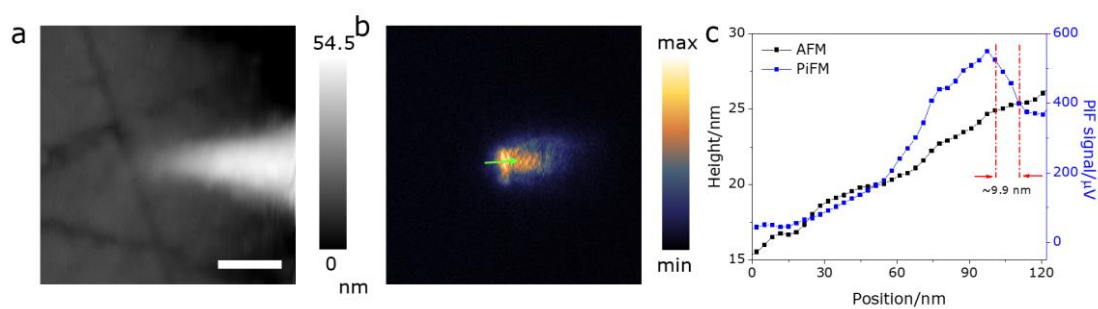

**Figure 4**. **PiFM measurements with high spatial resolution.** (a) AFM image of the tip end of an antenna fabricated with 6 μm PS sphere recorded in water. Scale bar: 200 nm. (b) PiF image at 1100 cm$^{-1}$ of the same region recorded simultaneously with (a). (c) Signals along the green arrow of topography (black curve) and PiFM images (blue curve).

A detailed image on the tip end of an antenna fabricated with 6 μm PS sphere was recorded in water to verify the spatial resolution of our platform (Figure 4a and 4b). As indicated in the simulation, the tip end of antenna has the highest electric filed enhancement. Since sharp geometric stage cannot be created by the antenna fabrication method, we define the spatial resolution as the minimum spatial distance that could clearly distinct the tip end of antenna. As we mentioned before, utilization of PDMS coated on AFM tip as probe could eliminate the sample artifact and the antenna end has a maximum optical response, we believe that the distance within distinguished electric field variation can be defined as the spatial resolution in the present work. According to the trace line shown in Figure 4c, the tip end with the highest PiF signal can be obviously distinguished, while the antenna height still gradually increases. Thus, the spatial resolution of our platform can be estimated as around 9.9 nm for a clear optical signal stage, which is very comparable with the PiFM detection of monolayers in air.[6] Since the thickness of PDMS layer on AFM tip is ~1-2 nm, the detection volume is calculated as around $7.69\times10^{-23}$ to $1.54\times10^{-22}$ L underneath the AFM tip within the spatial resolution of 9.9 nm. Assuming that PDMS has the similar density as the bulk materials (965 g/L), the monomer unit of PDMS (74 g/mol) can be calculated as ~604-1208 at sub-zeptomole level, which is 2 orders of magnitude higher than the best existing PTIR results[13] and is comparable to the molecular monolayer detection limit of PiFM in air.[6]

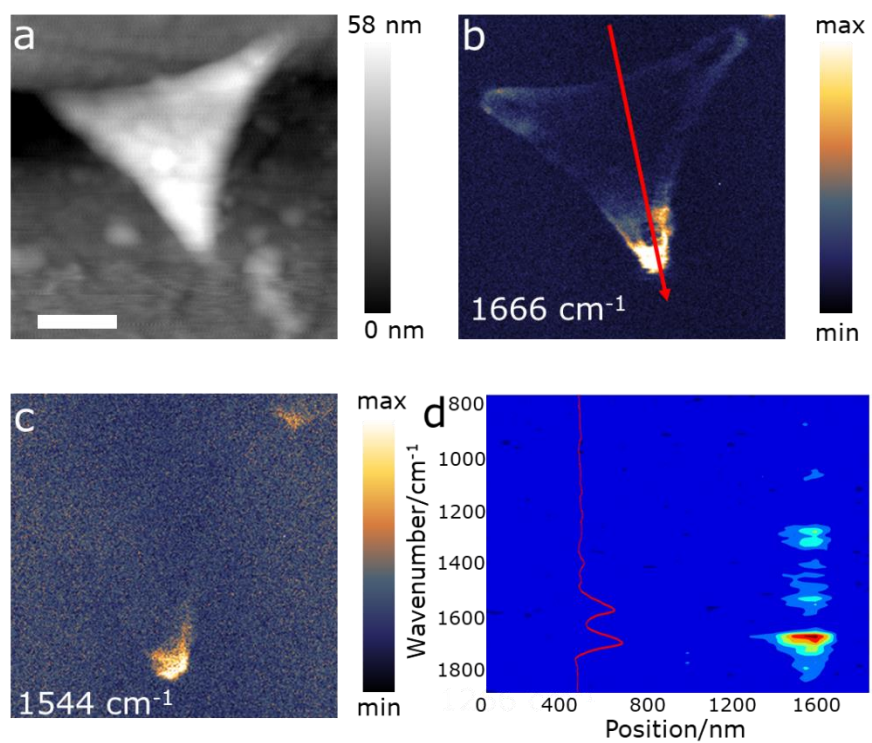

**Figure 5. PiFM measurements of bovine hemoglobin monolayer on an antenna fabricated with 4 μm PS sphere in water.** (a) AFM image of the antenna. Scale bar: 500 nm. (b) and (c): PiF images of the same antenna recorded at 1666 cm$^{-1}$ and 1544 cm$^{-1}$, respectively. (d) Corresponding spatio-spectral transection along the red arrows. Spectra were obtained in 5 s with 1 cm$^{-1}$ spectral resolution, with 50 nm separation.

The antenna enhanced strategy can also be used to sensitively monitor the samples covered on antenna surface. As demonstration, a monolayer of BHb, an important protein in respiratory chain of animals, was selected as the sample. Figure 5a shows the topography of an antenna fabricated with 4 μm PS sphere after the formation of BHb monolayer. The AFM image indicates that after BHb modification, the triangular shape of antenna is preserved. The height of the BHb modified antenna is around 50 nm, which is slightly higher than the pristine antenna, suggesting that only a monolayer of BHb is formed on the antenna surface since the size of BHb is known as 6×5×5 nm.[29]

Figure 5b and 5c display the PiF images recorded at 1666 cm$^{-1}$ and 1544 cm$^{-1}$, respectively assigning to the amide I and II bands of BHb.[30, 31] The recorded PiF images also exhibit clear position dependent signals, and the highest PiF signal appears at the antenna tip end. To figure out the origin of the position dependent PiF images, the PiF spectra were collected along the red arrows shown in Figure 5c. The spatio-spectral transection image (Figure 5d) indicates that at the antenna tip end (position ~1500 nm), two peaks centered at 1668 cm$^{-1}$ and 1548 cm$^{-1}$ are observed, while the PDMS peaks at 1266 cm$^{-1}$ and 1100 cm$^{-1}$ are also observable (PiF images recorded at 1266 cm$^{-1}$ is shown as Figure S10). The collected PiF spectra are in good accordance with the far field FTIR measurements as indicated by the red line in Figure 5d (direct spectral comparison shown in Figure S11). These results reveal that the PiF signal of BHb monolayer modified on antenna tip end can be successfully recorded, demonstrating the feasibility of our proposed strategy for detection of samples on the substrates. The present monolayer level sensitivity in aqueous environment is comparable to the recently reported results using scattering type scanning near field microscopy.[32]

**Conclusion**

In summary, we have proposed a novel nanoscale IR spectra and imaging method in aqueous environment based on the combination of PiFM and a mid-IR resonant antenna array. Our approach allows hundreds of molecules level detection and sub-10 nm spatial resolution in aqueous environment, which meets the sensitivity and resolution requirements for interfacial studies. With the enhanced PiF signal of PDMS modified on AFM tip, electric field distribution on the antenna can be clearly imaged, while with the resonance-dependent PiF signal, the chemical identification of PDMS on AFM tip and BHb monolayer on antenna has been successfully demonstrated. This work offers a sensitive force-based nanoscale IR technique in aqueous environment, which is expected to be broadly used in interfacial studies of chemical, biological, and electrochemical processes.

# Supporting Information for

# Antenna enhanced infrared photoinduced force imaging in aqueous environment with super-resolution and hypersensitivity

Jian Li,[1] Jie Pang,[1] Zhen-dong Yan,[2] Junghoon Jahng,[3] Jin Li,[1] William Morrison,[4] Jing Liang,[1] Qing-Ying Zhang,[1] Xing-Hua Xia[1*]

1. State key laboratory of analytical chemistry for life science, school of chemistry and chemical engineering, Nanjing University, Nanjing 210023
2. College of science, Nanjing Forestry University, Nanjing 210037
3. Center for nanocharacterization, Korea Research Institute of Standards and Science, Daejeon 34113
4. Molecular Vista Inc., 6840 Via Del Oro, Suite 110, San Jose, CA 95119
*Corresponding Author(s): Xing-Hua Xia: xhxia@nju.edu.cn

**Table of content**

S1. Measured antenna resonance spectra (normalized) in air.
S2. Simulated antenna resonance spectra in water of an antenna array with different incident angles.
S3. Simulated $|E_x/E_0|^2$ distribution on the antenna fabricated with 5 μm PS sphere.
S4. Dielectric function of PDMS.
S5. PiF images on the antenna fabricated with 5 μm PS sphere at different wavenumbers.
S6. Comparison of the PiF signal distribution and simulated $E_z$ distribution.
S7. PiFM measurements on the antenna fabricated with 4 μm PS sphere.
S8. PiFM measurements on the antenna fabricated with 6 μm PS sphere.
S9. PiF images on antenna fabricated with 6 μm PS sphere at different wavenumbers.
S10. PiF image of BHG monolayer on an antenna fabricated with 4 μm PS sphere recorded at 1266 cm$^{-1}$ and 1544 cm$^{-1}$.
S11. Comparison of PiF spectrum and ATR-SEIRAS spectrum of BHb.

**S1. Measured antenna resonance spectra (normalized) in air.**

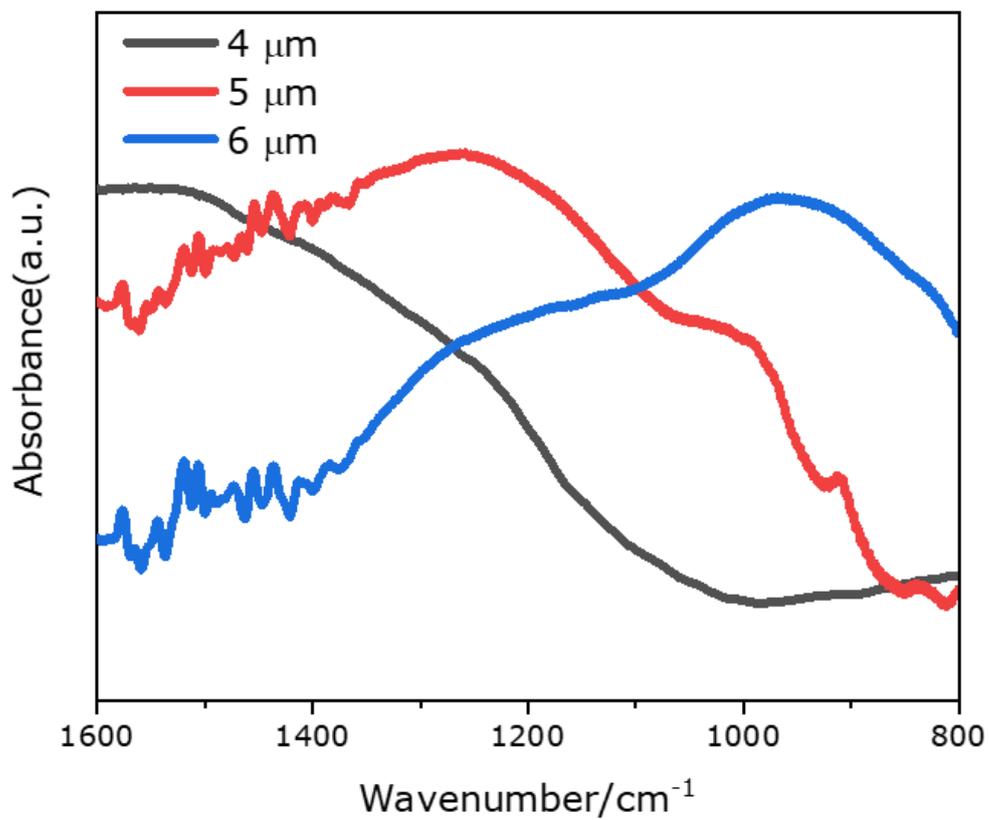

Figure S1. Measured antenna resonance spectra (normalized) in air of the antenna array fabricated with PS spheres with diameters of 4, 5 and 6 μm, respectively. (Substrate: ZnSe of n=2.4, incident angle: 0 degree)

**S2. Simulated antenna resonance spectra in water of an antenna array with different incident angles.**

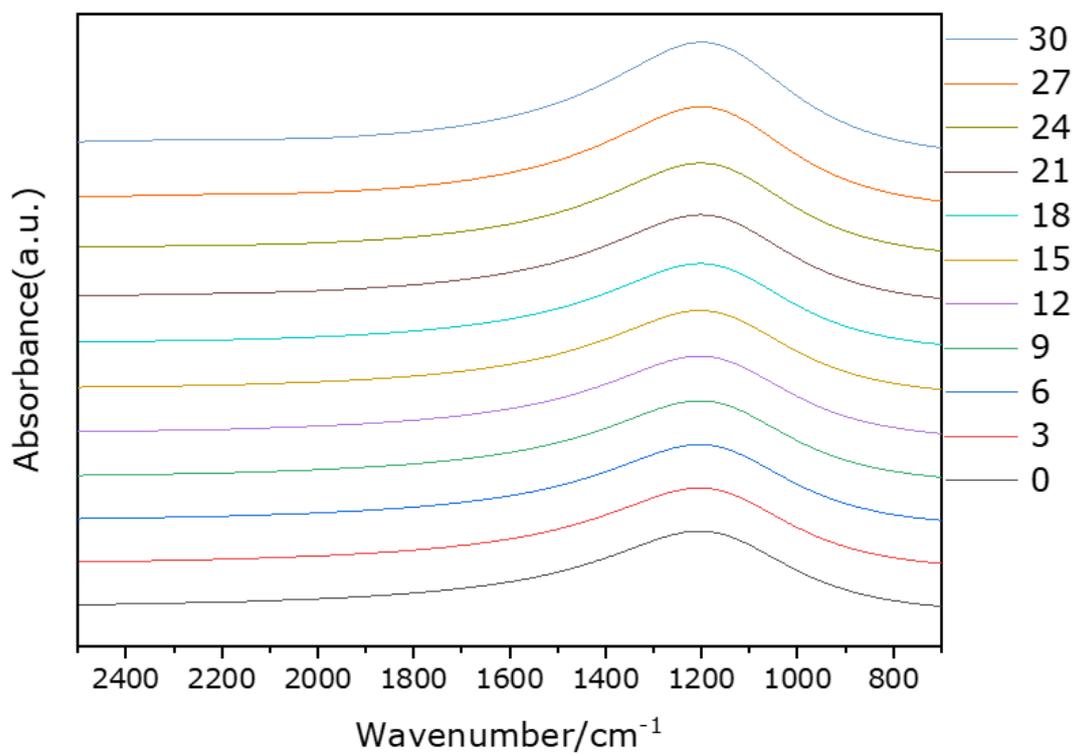

Figure S2. Simulated antenna resonance spectra (normalized) in water of an antenna array with different incident angles (Substrate: ZnSe of n=2.4, mask size: 5 µm).

**S3. Simulated |Ex/E₀|² distribution on the antenna fabricated with 5 μm PS sphere.**

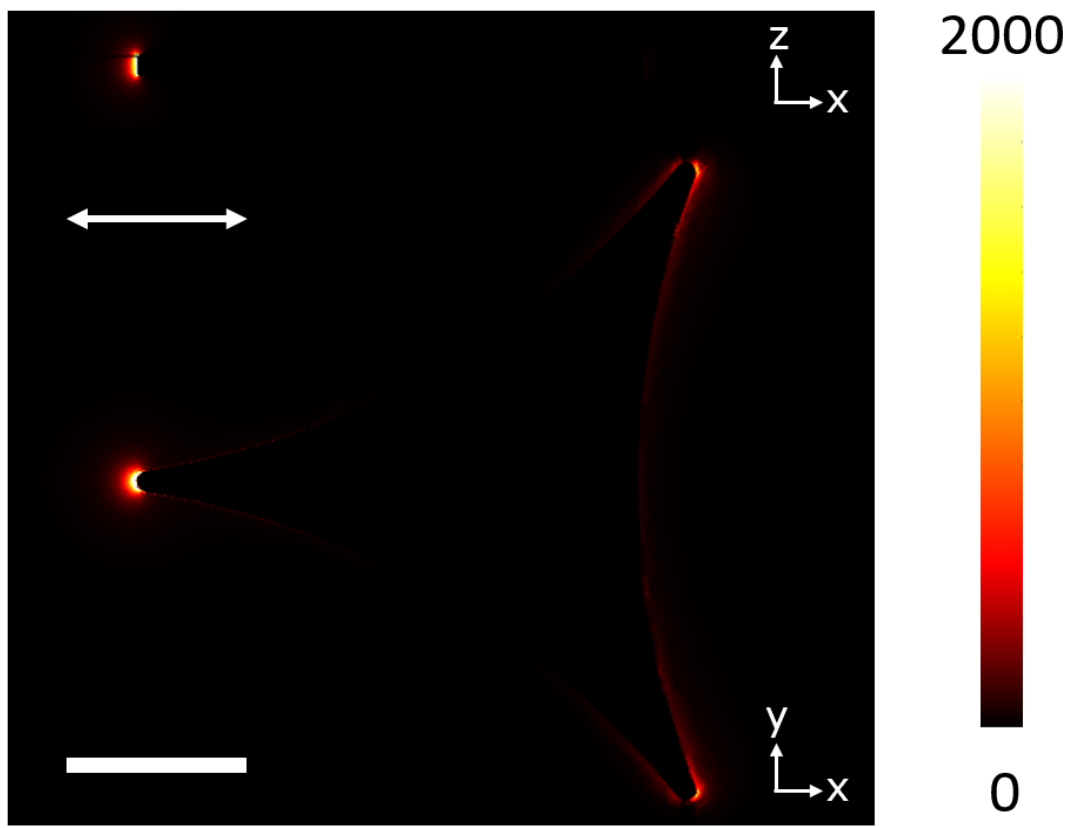

Figure S3. Simulated $|E_x/E_0|^2$ distribution on antenna fabricated with 5 μm PS sphere. Wavenumber: 1205 cm$^{-1}$; xy plane: 1 nm away in water from the antenna surface; xz plane: along the main axe of the antenna; the incident light is with *s* polarization and the direction is indicated by the double-head arrow with incident angel of 12 degree; scale bar in xy plane: 500 nm.

**S4. Dielectric function of PDMS.**

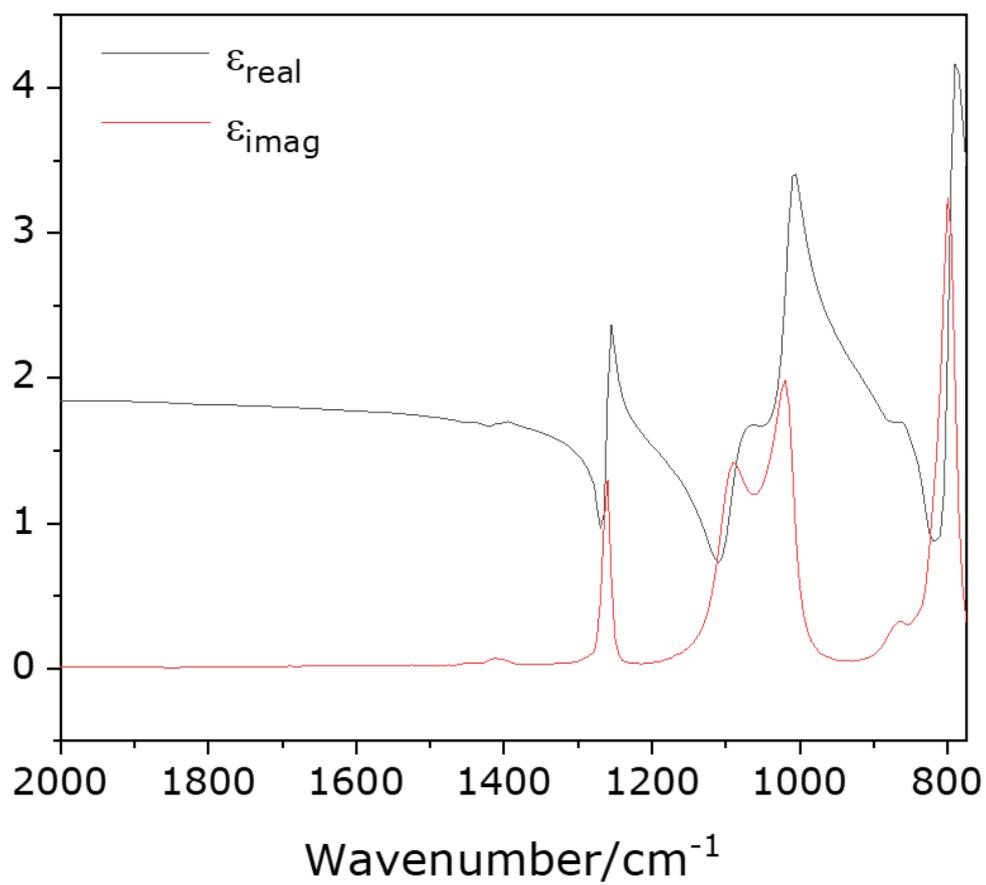

Figure S4. Dielectric function of PDMS in the detection range. The refractive index data are from Ref. 2.

**S5. PiF images on the antenna fabricated with 5 µm PS sphere at different wavenumbers.**

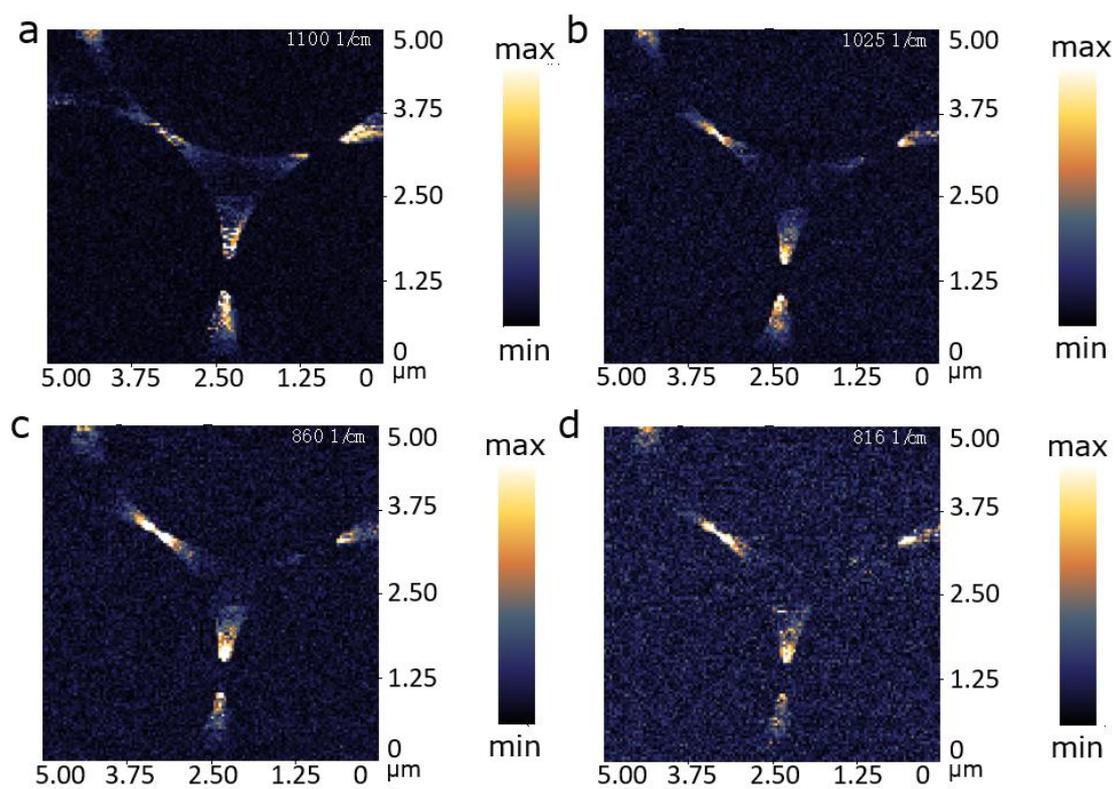

Figure S5. PiF images of the same region recorded at (a) 1100 cm$^{-1}$, (b) 1025 cm$^{-1}$, (c) 860 cm$^{-1}$ and (d) 816 cm$^{-1}$, respectively. The antenna was fabricated with 5 µm PS sphere.

## S6. Comparison of the PiF signal distribution and simulated $E_z$ distribution.

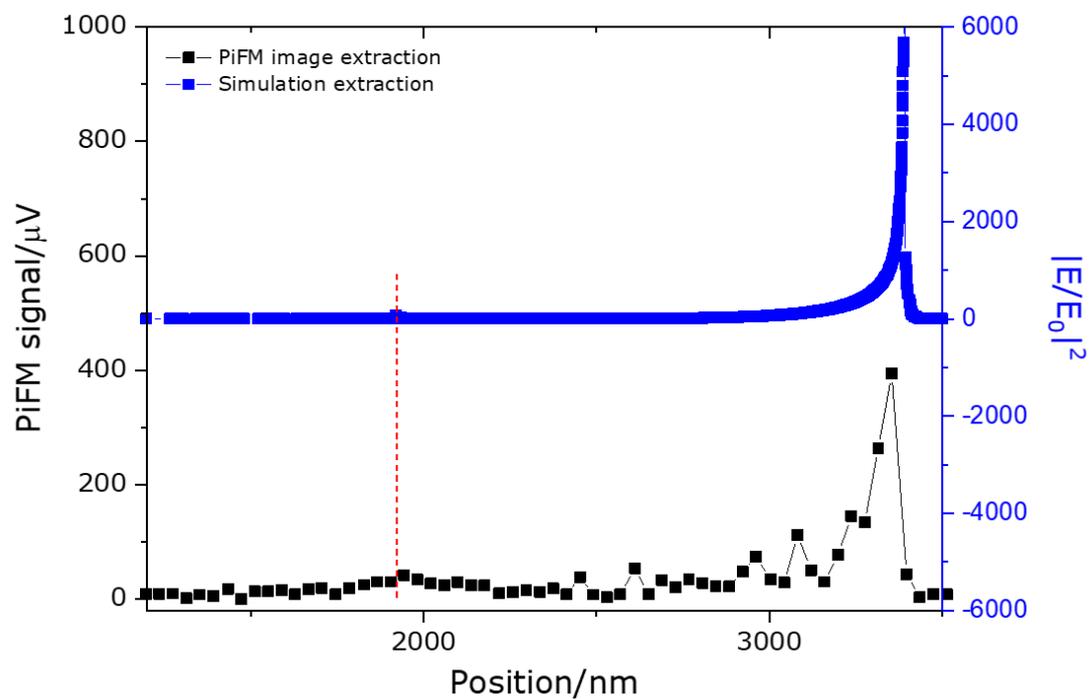

Figure S6. Comparison of the PiF signal distribution and simulated $E_z$ distribution along the main axis and on top of the triangle antenna. The antenna was fabricated with 5 µm PS sphere.

**S7. PiFM measurements on the antenna fabricated with 4 μm PS sphere.**

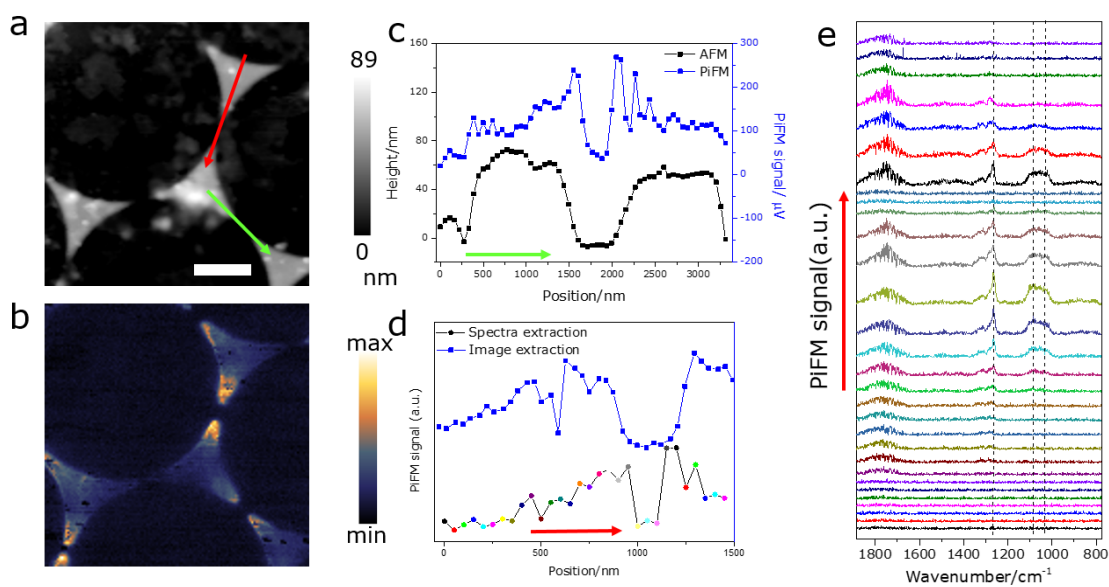

**Figure S7. PiFM measurements on the antenna fabricated with 4 μm PS sphere.** (a) AFM image recorded in water. Scale bar: 1 μm; (b) PiF image of the same region recorded at 1750 cm$^{-1}$ simultaneously with (a); (c) Signal along the red arrows of topography (black curve) and PiFM image (blue curve). d) Extracted data point at 1750 cm$^{-1}$ from the PiFM spectra recorded at different points along the green arrow in (a) with 100 nm separation (colored curve) and the PiFM signal along the green arrow of PiFM image (blue curve). (e) PiFM spectra recorded on the position colored-indicated in (d).

In Figure S7-S9, the wavenumber selected for image are based on two considerations:
(1) At the IR absorption peak position of PDMS.
Wavenumber: 1266 cm$^{-1}$, 1100 cm$^{-1}$, 1025 cm$^{-1}$ and 816 cm$^{-1}$.
(2) At the highest power density of the QCL laser.
Wavenumber: 1750 cm$^{-1}$ and 869 cm$^{-1}$.

**S8. PiFM measurements on the antenna fabricated with 6 μm PS sphere.**

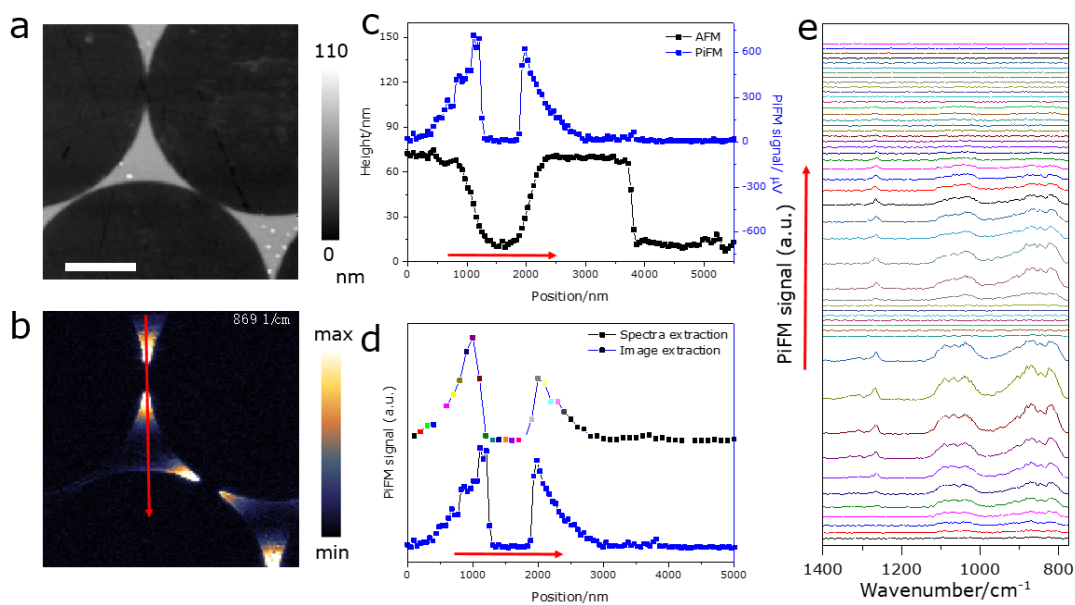

**Figure S8. PiFM measurements on the antenna fabricated with 6 μm PS sphere.** (a) AFM image recorded in water. Scale bar: 2 μm (b) PiF images of the same region recorded at 869 cm$^{-1}$ simultaneously with (a). (c) Signal along the red arrow of topography (black curve) and PiFM images (blue curve). d) Extracted data point at 869cm$^{-1}$ from the PiFM spectra recorded at different points along the red arrow in (a) with 100 nm separation (colored curve) and the PiFM signal along the red arrow of PiFM image (blue curve). (e) PiFM spectra recorded on the position colored-indicated in (d).

**S9. PiF images on antenna fabricated with 6 µm PS sphere at different wavenumbers.**

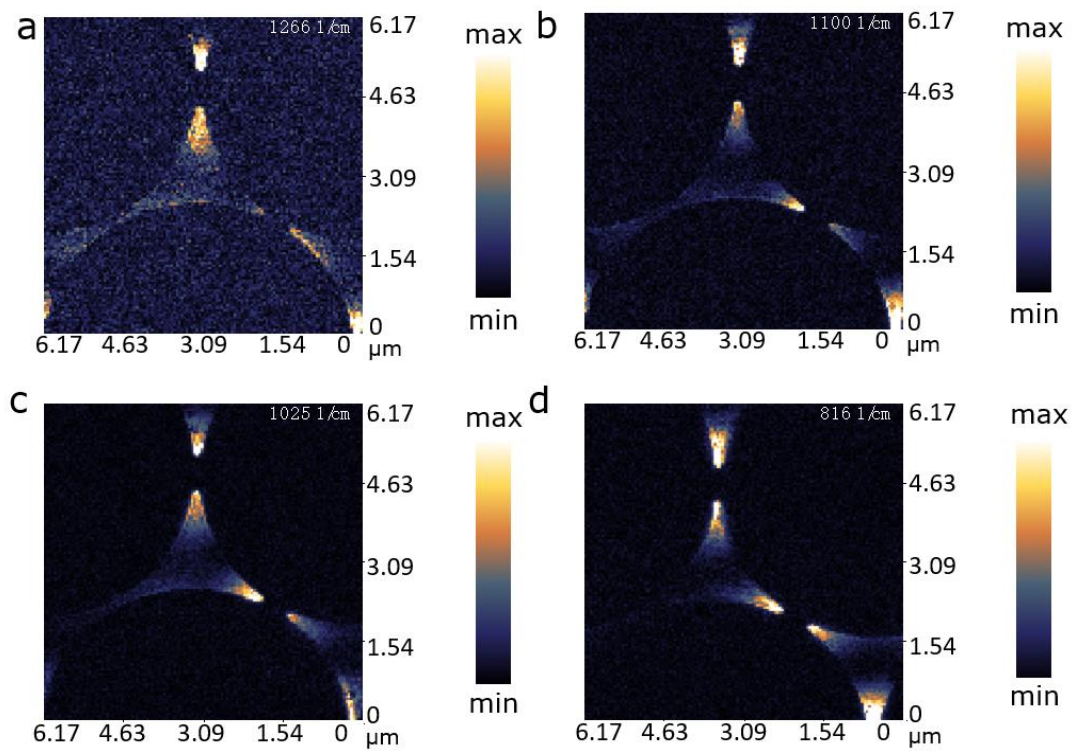

Figure S9. PiF images of the same region recorded at (a) 1266 cm$^{-1}$, (b) 1100 cm$^{-1}$, (c) 1025 cm$^{-1}$ and (d) 816 cm$^{-1}$, respectively. The antenna was fabricated with 6 µm PS sphere.

**S10. PiF image of BHb monolayer on an antenna fabricated with 4 µm PS sphere recorded at 1266 cm$^{-1}$.**

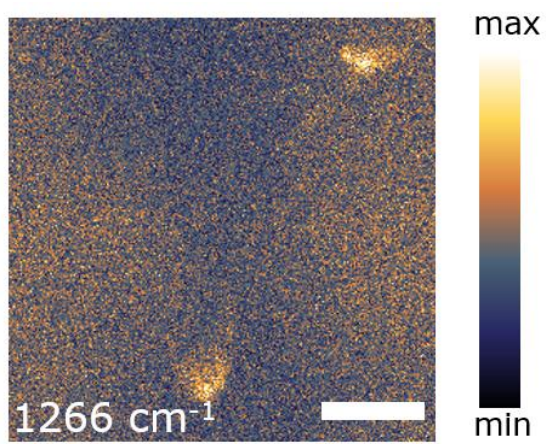

**Figure S10.** PiF image of BHb monolayer on the same antenna as in Figure 5 recorded at 1266 cm$^{-1}$. Scale bar: 500 nm.

## S11. Comparison of PiF spectrum and ATR-SEIRAS spectrum of BHb.

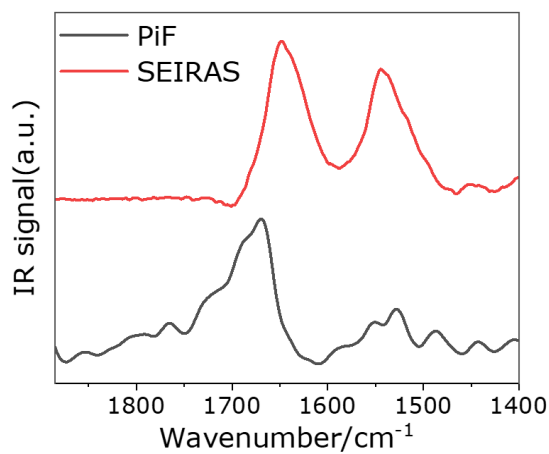

Figure S11. PiF (black curve) and ATR-SEIRAS (red curve) spectra of BHb on gold antenna and gold nanofilm, respectively.

The ATR-SEIRAS spectrum of BHb was recorded on a Nicolet IS 50 with a homemade ATR accessory. Unpolarized IR radiation was totally reflected at the Au nano film/solution interface with an incident angle θ=75° and was detected with a liquid-nitrogen-cooled MCT detector. The Au nanofilm/ZnSe prism as the ATR-SEIRAS enhanced substrate was fabricated by an electroless deposition method developed in our group.[1] After adding BHb solution on the Au nanofilm surface for 30 min, the spectrum of BHb was collected by taking the IR spectrum of water as the reference.

References:
[1] Bao, W. J.; Li, J.; Li, J.; Zhang, Q. W.; Liu, Y.; Shi, C. F.; Xia, X. H. Au/ZnSe-based surface enhanced infrared absorption spectroscopy as a universal platform for bioanalysis. *Anal. Chem.,* **2018,** 90, 3842-3848.